\documentclass[pdflatex,sn-mathphys-num]{sn-jnl}

\usepackage{graphicx}%
\usepackage{multirow}%
\usepackage{amsmath,amssymb,amsfonts}%
\usepackage{amsthm}%
\usepackage{mathrsfs}%
\usepackage[title]{appendix}%
\usepackage{xcolor}%
\usepackage{textcomp}%
\usepackage{manyfoot}%
\usepackage{booktabs}%
\usepackage{algorithm}%
\usepackage{algorithmicx}%
\usepackage{algpseudocode}%
\usepackage{listings}%
\usepackage{caption}
\usepackage{subcaption}

\theoremstyle{thmstyleone}%
\usepackage{algpseudocode}

\theoremstyle{thmstyletwo}%

\theoremstyle{thmstylethree}%

\usepackage[defaultcolor=red]{changes}  
\definechangesauthor[name={Long}, color=blue]{A} 

\presetkeys{todonotes}{color=yellow!30,bordercolor=red,fancyline}{} 

\begin{document}
\title[Article Title]{
CFDagent: A Language-Guided, Zero-Shot Multi-Agent System for Complex Flow Simulation}
\author[1,2]{\fnm{Zhaoyue} \sur{Xu}}\email{zhaoyue@chalmers.se}
\author[1,3]{\fnm{Long} \sur{Wang}}\email{wanglong@imech.ac.cn}
\author[1,3]{\fnm{Chunyu} \sur{Wang}}\email{wangchunyu@imech.ac.cn}
\author[1,3]{\fnm{Yixin} \sur{Chen}}\email{chenyixin@imech.ac.cn}
\author[1,3]{\fnm{Qingyong} \sur{Luo}}\email{luoqingyong@imech.ac.cn}
\author*[1,2]{\fnm{Hua-Dong} \sur{Yao}}\email{huadong.yao@chalmers.se}
\author*[1,3]{\fnm{Shizhao} \sur{Wang}}\email{wangsz@lnm.imech.ac.cn}
\author[1,3]{\fnm{Guowei} \sur{He}}\email{hgw@lnm.imech.ac.cn}

\affil[1]{\orgdiv{The State Key Laboratory of Nonlinear Mechanics}, \orgname{Institute of Mechanics, Chinese Academy of Sciences}, \orgaddress{\city{Beijing}, \postcode{100190}, \country{China}}}

\affil[2]{\orgdiv{Department of Mechanics and Maritime Sciences}, \orgname{Chalmers University of Technology}, \orgaddress{\city{Gothenburg}, \postcode{41296}, \country{Sweden}}}

\affil[3]{\orgdiv{School of Engineering Sciences}, \orgname{University of Chinese Academy of Sciences}, \orgaddress{\city{Beijing}, \postcode{100049}, \country{China}}}

\abstract{
We introduce CFDagent, a zero-shot, multi-agent system that enables fully autonomous computational fluid dynamics (CFD) simulations from natural language prompts. CFDagent integrates three specialized LLM-driven agents: (i) the Preprocessing Agent that generates 3D geometries from textual or visual inputs using a hybrid text-to-3D diffusion model (Point-E) and automatically meshes the geometries; (ii) the Solver Agent that configures and executes an immersed boundary flow solver; and (iii) the Postprocessing Agent that analyzes and visualizes the results, including multimodal renderings. These agents are interactively guided by GPT-4o via conversational prompts, enabling intuitive and user-friendly interaction. We validate CFDagent by reproducing canonical sphere flows at Reynolds numbers of 100 and 300 using three distinct inputs: a simple text prompt (i.e., “sphere”), an image-based input, and a standard sphere model. The computed drag and lift coefficients from meshes produced by each input approach closely match available data. The proposed system enables synthesization of flow simulations and photorealistic visualizations for complex geometries. Through extensive tests on canonical and realistic scenarios, we demonstrate the robustness, versatility, and practical applicability of CFDagent. By bridging generative AI with high-fidelity simulations, CFDagent significantly lowers barriers to expert-level CFD, unlocking broad opportunities in education, scientific research, and practical engineering applications.
}




\maketitle

\section{Main}\label{sec1}
Large Language Models (LLMs), particularly OpenAI's GPT series~\citep{Radford2018ImprovingLU,radford2019language,NEURIPS2020,GPT4}— have significantly expanded the capabilities of artificial intelligence, driving transformative advances in natural language processing, reasoning, and zero-shot learning.
Rooted fundamentally in the Transformer architecture~\citep{Vaswani2017}, these models have revolutionized diverse fields by enhancing tasks related to text generation, knowledge extraction, and generative functionalities~\citep{GPT4,NEURIPS2020}.
Beyond natural language applications, GPT models also exhibit strong capabilities in code generation and computational physics, achieving promising results in various programming-related benchmarks~\citep{GPT4,Jacob2021}.

Despite their impressive capabilities, current standalone LLMs remain limited in tasks requiring extensive, precise, and systematic content generation—particularly in computational physics and simulation~\citep{li2023camel} due to persistent challenges such as hallucination~\citep{farquhar2024detecting}.
As elucidated by Wolfram~\citep{wolfram2024can}, fundamental constraints inherent to these models restrict their ability to autonomously ``solve science".
In particular, although LLMs can effectively assist in accelerating the development of numerical solvers in fluid dynamics or other physical problems~\citep{Jacob2021}, they remain insufficient as replacements for simulations, especially in tasks involving complex geometry handling, rigorous physical modeling, and sophisticated numerical methods such as Computational Fluid Dynamics (CFD). 
Despite significant research integrating AI into CFD workflows~\citep{gao2021phygeonet,du2024conditional,duraisamy2019turbulence,brunton2020machine,kashefi2025kolmogorov,yuan2025dimensionless}, current AI technologies—including LLMs—nevertheless remain unable to function as direct solvers for the underlying governing equations~\citep{wolfram2024can}.

To overcome these limitations, integrating LLM capabilities into simulation frameworks has emerged as an increasingly promising direction.
Recent studies~\citep{zhang2024usinglargelanguagemodels,chen2024metaopenfoamllmbasedmultiagentframework,Pandey2025,Du2024} have proposed hybrid approaches leveraging LLM-driven techniques for enhancing simulations and engineering design optimization.
Notably, these approaches commonly involve interactive natural language prompts, automatic equation formulation, and assisted simulation frameworks.
For example, MetaOpenFOAM~\citep{chen2024metaopenfoamllmbasedmultiagentframework} proposes an automated CFD framework based on multi-agent systems. 
Similarly, OpenFOAM-GPT~\citep{Pandey2025} introduces a RAG-augmented LLM framework tailored to OpenFOAM-based CFD simulations.
Nevertheless, existing OpenFOAM-based CFD agent systems continue to rely on manually curated case libraries for knowledge retrieval, potentially limiting their scalability and generalizability to previously unseen or highly complex flow scenarios.
Consequently, such systems often require substantial supervision when dealing with atypical or high-complexity tasks, thereby limiting their degree of end-to-end automation.
This limitation highlights that current frameworks are not yet capable of robust zero-shot generalization and still depend on task-specific adaptation or expert intervention to operate effectively in novel problem domains.

\begin{figure}[h]
\centering
\includegraphics[width=1.0\textwidth]{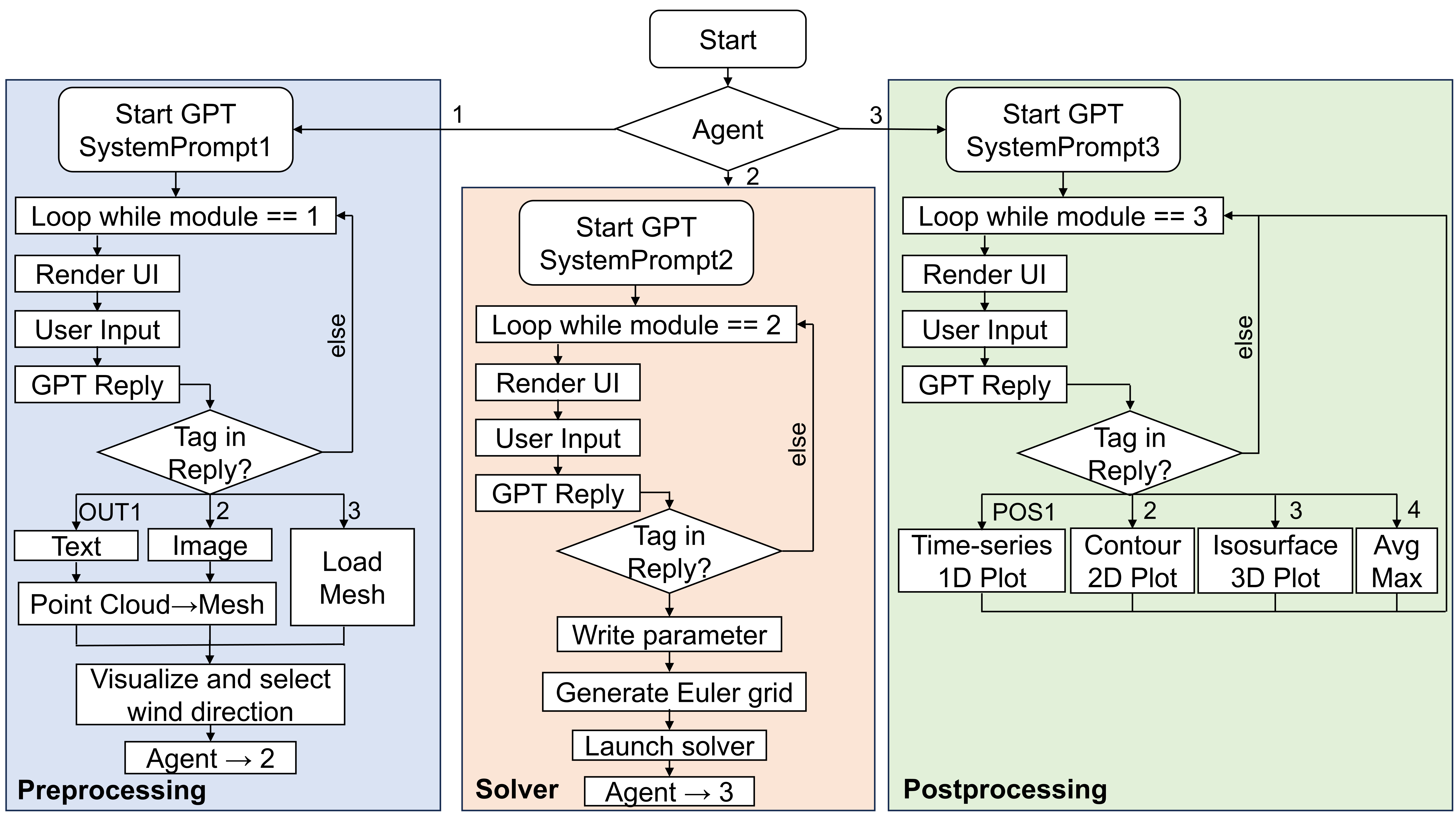}
\caption{\textbf{Workflow diagram of CFDagent.} The system consists of three core agents: the Preprocessing Agent, the Solver Agent, and the Postprocessing Agent. Each agent's role is defined through a tailored system prompt that conditions the LLM to perform stage-specific tasks within the simulation workflow. The agents interact with the user via natural language prompts, enabling dynamic input interpretation and autonomous execution of key operations. These include geometry and mesh generation, solver configuration and flow simulation, as well as postprocessing for quantitative analysis and result visualization.}\label{fig:Workflow}
\end{figure}

In this work, we propose an LLM-driven multi-agent framework, CFDagent, to enable fully language guided, zero-shot and end-to-end complex flow simulations.
Through natural language interaction, the framework is designed to manage three core stages that constitute the workflow of a typical complex flow simulation: 1) preprocessing, which includes geometry and mesh generation; 2) flow solving, which involves numerically solving the governing fluid dynamics equations under specified initial and boundary conditions; and 3) postprocessing, which encompasses the analysis and visualization of simulation results.
Correspondingly, CFDagent leverages LLM-driven agents to overcome the substantial demands on human and computational resources typically required in these stages, as illustrated in Fig.~\ref{fig:Workflow}. 

To reform the first stage, the Preprocessing Agent integrates a hybrid text-to-3D diffusion model, point-E, to autonomously generate geometry and mesh.
Point-E generates semantically accurate 3D point clouds from natural-language prompts by first synthesizing images via GLIDE~\citep{nichol2021improveddenoisingdiffusionprobabilistic} and then reconstructing geometry, achieving 10-100x faster sampling than earlier text-to-3D approaches~\citep{Jain_2022_CVPR,Lin_2023_CVPR,sanghi2022clip,mittal2023autosdfshapepriors3d,FuNEURIPS2022_3a33ae4d,zeng2022lionlatentpointdiffusion,Sanghi2023,nichol2022pointegenerating3dpoint}.
In the second stage, the Solver Agent enables the LLM to guide users in configuring simulation parameters and launching the flow solver.
Complementing this, our Immersed Boundary (IB) method, rooted in Peskin's foundational work~\citep{PESKIN1972252,Peskin_2002,Mittal2005} and leveraging direct-forcing methods~\citep{mohd1997simulations,FADLUN200035,Iaccarino2003,KIM2001132,YANG200612,ZHANG2007250,UHLMANN2005448,TAIRA20072118,YANG20097821}, imposes boundary conditions on complex geometries without body-fitted meshes by solving small, localized linear systems~\citep{WANG20113479}.
Enhanced through parallel domain decomposition, our implementation delivers both high accuracy and computational efficiency, as validated by extensive benchmarks~\citep{WANG20113479,Chen_Liu_Wang_2025,Wang2023Bayesian,Zhou2022Wall,Wang2021}.
Lastly, the Postprocessing Agent equips the LLM with scripts for analyzing physical quantities, visualizing flow fields, and fusing simulation results to generate realistic imagery of physical objects.

\section{Results and discussion}\label{Results}
\subsection{Autonomous flow simulation around a sphere as a canonical geometry}
By leveraging the capabilities of the proposed LLM-driven CFD framework, we conducted three-dimensional numerical simulations of unsteady flow around a sphere at Reynolds numbers of 100 and 300.
The sphere is a canonical geometry that serves as a benchmark for validating the accuracy and reliability of the simulation framework.

\begin{table}[htbp]
\centering
\renewcommand{\arraystretch}{1.3}
\begin{tabular}{p{3cm}p{8cm}}
\hline
\textbf{User Prompt} & \textbf{CFDagent} \\
\hline
Compute the flow around a sphere at Re300
 & We have the object as ``sphere" and the Reynolds number as ``300". Please confirm these before I initiate the simulation. \\
Yes, I confirm. &  Starting point cloud generation for ``sphere". Please wait... Creating SDF and generating mesh...  Please select a wind direction. \\
\texttt{[Click a direction]} &  What is the end time of the simulation? \\
200 should be enough.&  Great! You've set the end time of the simulation to 200.0.
Now, what is the CFL number? (The default is 0.8)\\
Use default.&  You've set the CFL number to 0.8.
Now, what is the maximum time step? (The default is 0.1)\\
I prefer 0.2. &  You've set the maximum time step to 0.2.
Finally, what is the output interval time? (The default is 1.0)\\
10&  You've set the output interval time to 10.0.
\textit{Here's a summary of all values.}
Do you confirm these values?\\
Yes!&  Use the default Euler grid or customize it?\\
Default.&  Parameters saved to Para.txt and Bc\_Ic.txt. Euler grid is generated with default settings. Simulation has started!\\
\hline
\end{tabular}
\caption{An example interaction between the user and CFDagent during a simulation of flow around a sphere. The complete dialogue example can be found in Supplementary Information.}
\label{tab:cfd-dialogue}
\end{table}

As shown in Table~\ref{tab:cfd-dialogue}, when the user expresses interest in CFD, CFDagent autonomously orchestrates its core agents to initiate the procedure.
The Preprocessing Agent requests the geometry and Reynolds number, supporting three distinct input modes: (1) direct generation from a natural language prompt (e.g., ``sphere"), (2) reconstruction from an input image of the object, and (3) direct import of a pre-generated mesh.
All three modes ultimately produce a Lagrangian surface mesh representing the object.
Subsequently, the Solver Agent prompts the user for detailed numerical simulation settings and initiates the solver.
Details of the default Eulerian grid used in this study are provided in the Supplementary Information.
Mesh convergence was rigorously validated through multiple tests, confirming that variations in mesh resolution and domain size have a negligible effect on the flow characteristics~\citep{WANG20113479,WANG2013210,Wang2021,Chen_Liu_Wang_2025}.

The Postprocessing Agent generates figures to visualize the simulation results as specified by the user prompt. In the post-processing step, we compute the dimensionless drag and lift coefficients, $C_d$ and $C_l$, by normalizing the drag ($D$) and lift ($L$) forces by the dynamic pressure by the dynamic pressure, $q =\frac{1}{2} \rho U^2$, and a reference area, $S$:  
\begin{equation}
C_d = D/(q S), C_l = L/(q S).  
\end{equation}
Here, $\rho$ is the fluid density and $U$ the freestream velocity.


At Re$=100$, the flow around a sphere is nearly steady and axisymmetric, with a well-defined separation bubble forming downstream.
The lift coefficient $C_l$ under this condition is nearly zero, consistent with the axisymmetric nature of the flow.
Table~\ref{tab:drag_comparison_Re100} provides a comparison of the drag coeffient $C_d$ from numerical simulations employing different geometry-inputting approaches, as well as previously reported literature values.
The comparison indicates good agreement, thus validating the accuracy and reliability of the current approaches.

\begin{figure}[h]
\centering
\includegraphics[width=0.5\textwidth]{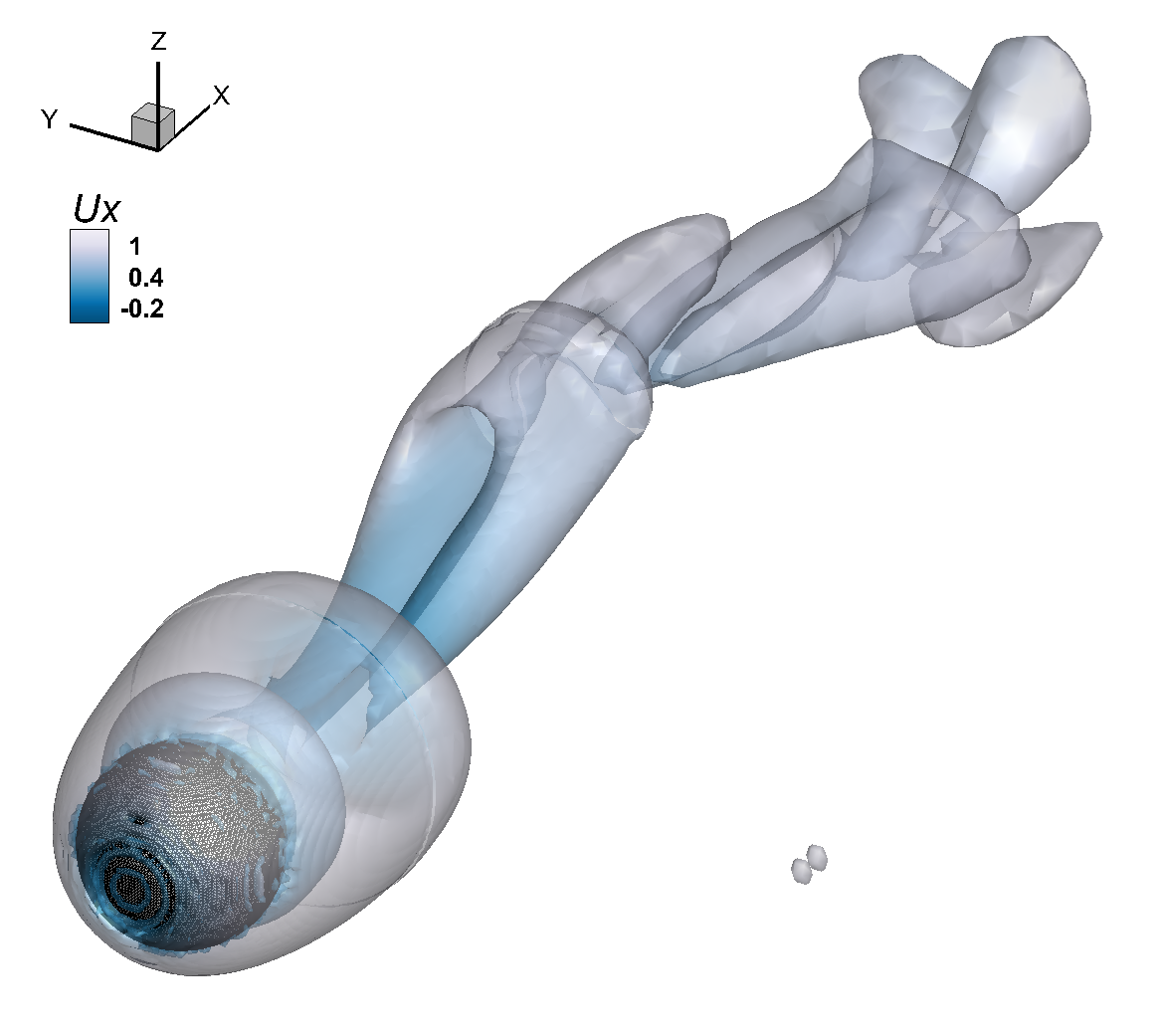}
\caption{\textbf{Instantaneous vortical structure for flow past a sphere at Re = 300.}
The vortical structures are visualized using iso-surfaces of the $Q$-criterion, highlighting regions with a high ratio of vorticity to strain. The iso-surfaces are colored by the streamwise velocity, ranging from -0.2 to 1, with colors transitioning from blue to white, illustrating the variation of the flow speed around the vortex structures.}\label{sphere}
\end{figure}

At Re$=300$, the flow transitions to an unsteady state characterized by pronounced vortex shedding and associated fluctuations in $C_d$ and $C_l$.
A representative instantaneous vortical structure under this condition is illustrated in Fig.~\ref{sphere}, where vortices are visualized and identified based on the $Q$-criterion, a common method for vortex identification.
The observed vortex shedding patterns are consistent with those previously documented by \citet{WANG20113479}.
Table~\ref{tab:drag_comparison_Re300} compares the computed time-averaged drag and lift coefficients against literature benchmarks, demonstrating excellent quantitative agreement and further supporting the robustness and accuracy of the simulation methodology employed.
\begin{table}[htbp]
    \centering
    \caption{Comparison of drag coefficients ($C_d$) of the sphere at Re$=100$.}
    \label{tab:drag_comparison_Re100}
    \begin{tabular}{ll}
        \toprule
        \textbf{Case} & $C_d$ \\
        \midrule
        Prompt-based geometry (``sphere") & 1.11 \\
        Image-based geometry & 1.11 \\
        Imported mesh geometry & 1.12 \\
        Johnson and Patel~\citep{JOHNSON_PATEL_1999} & 1.10 \\
        Fadlun et al.~\citep{FADLUN200035} & 1.08 \\
        \bottomrule
    \end{tabular}
\end{table}
\begin{table}[htbp]
    \centering
    \caption{Comparison of drag and lift coefficients ($C_d$ and $C_l$) of the sphere at Re$= 300$.}
    \label{tab:drag_comparison_Re300}
    \begin{tabular}{lll}
        \toprule
        \textbf{Case} & $C_d$ & $C_l$\\
        \midrule
        Prompt-based geometry (``sphere") & 0.68 & 0.065\\
        Image-based geometry & 0.68 & 0.065\\
        Imported mesh geometry & 0.68 & 0.071\\
        Johnson and Patel~\citep{JOHNSON_PATEL_1999} & 0.66 & 0.069\\
        Kim et al.~\citep{KIM2001132} & 0.66 & 0.067\\
        \bottomrule
    \end{tabular}
\end{table}

\subsection{Zero-shot simulation of complex flows over arbitrary real-world objects}
The Preprocessing Agent of the proposed LLM-driven CFD framework is capable of generating complex geometries based solely on natural language prompts, without requiring any pre-existing CAD models or case-specific training.
This capability allows for the simulation of arbitrary objects, significantly enhancing the versatility and applicability of the framework. 
Here, we select the geometry ``human" as an illustrative scenario to demonstrate the framework's end-to-end automated simulation pipeline.
Initially, the Preprocessing Agent receives the object description ``human" and the specified Reynolds number, ${\rm Re}=300$.
The agent automatically generates the corresponding geometric model and computational mesh using the integrated text-to-3D model and the previously described mesh conversion methods.
The simulation employs the same settings as those used in the sphere flow cases to ensure consistency across scenarios.
Consequently, the prompt is similar to that of Table~\ref{tab:cfd-dialogue}, differing only in the geometry name, and is therefore omitted.

\begin{figure}[h]
\centering
\includegraphics[width=0.9\textwidth]{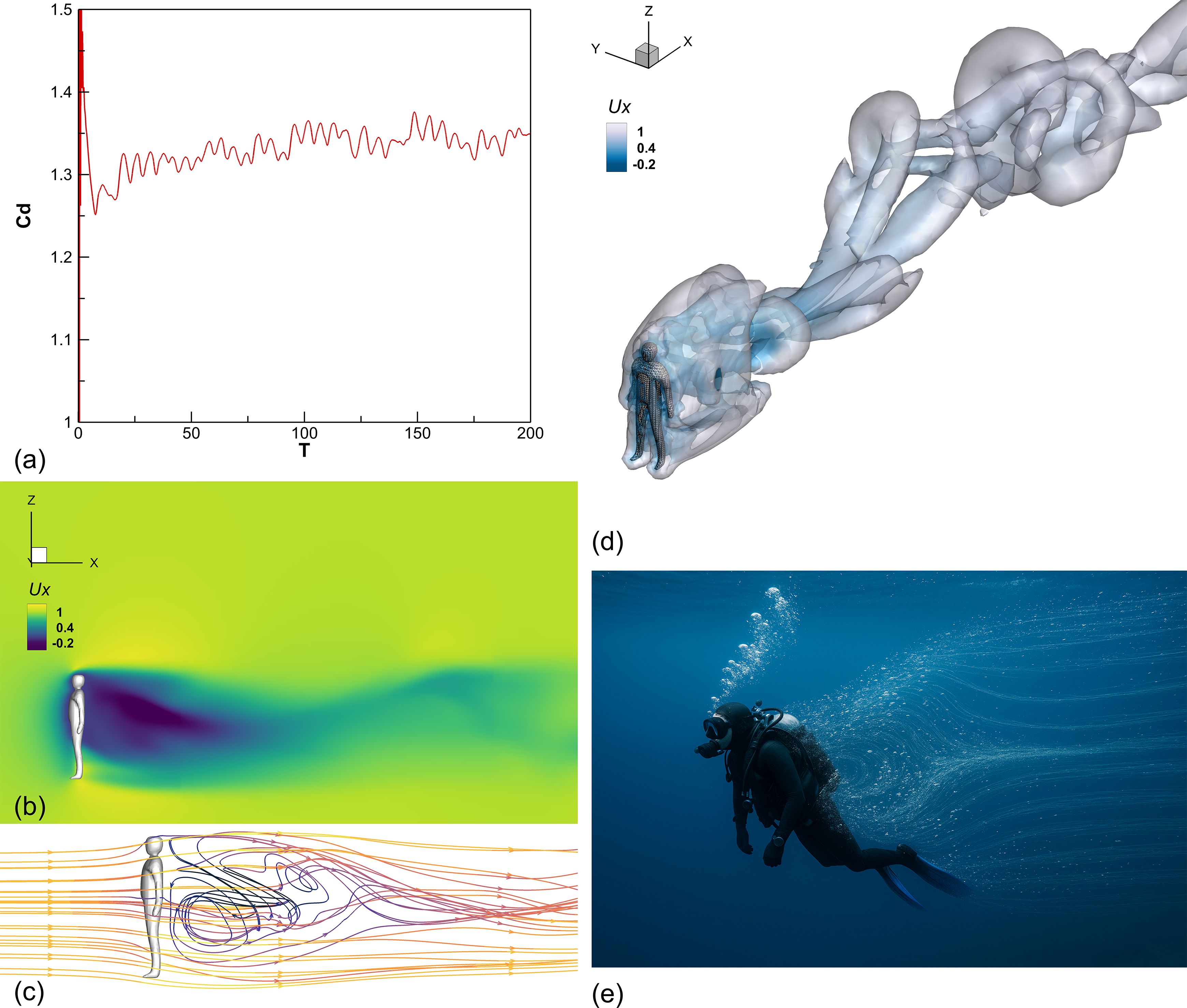}
\caption{\textbf{Postprocessing results from CFDagent for flow past a human geometry at Re$ = 300$.} (a) instantaneous drag coefficient, (b) streamwise velocity component at slice $y=0$, (c) streamlines colored by velocity magnitude at slice $y=0$, (d) iso-surface visualization of the $Q$-criterion ($Q=0.1$) colored by streamwise velocity, and (e) GPT-4 synthesized visualization based on simulation results.}
\label{man}
\end{figure}

After completing the simulation, the Postprocessing Agent executes several visualization and analysis commands provided through natural language prompts.
Issuing the prompt ``Plot the instantaneous drag coefficient over time." triggers the \texttt{POS1} analysis routine illustrated in Fig.~\ref{fig:Workflow}.
The instantaneous $C_d$ of the human geometry is extracted from the data file and subsequently plotted, as shown in Fig.~\ref{man}(a).
The time-averaged drag coefficient obtained from this simulation is approximately $C_d \approx 1.3$.
Subsequently, we explore additional flow features using natural language prompts.
Responding to the prompt ``Visualize the $u_x$ velocity component on the slice $y=0$", the agent generates the velocity distribution in the streamwise direction on the specified plane, as illustrated in Fig.~\ref{man}(b).
With the command ``Streamlines on the $y=0$ slice, colored by velocity magnitude with the cmocean colormap.", the flow streamlines on this slice are visualized in Fig.~\ref{man}(c).
These two figures clearly highlight the velocity field structure around the geometry.
Further, the instruction ``Generate the iso-surface of the $Q$-criterion at $0.1$, and color it according to the streamwise velocity $u_x$ using a blue-to-white colormap." results in the visualization of three-dimensional vortex structures, as depicted in Fig.~\ref{man}(d).
Finally, leveraging the image synthesis capabilities of GPT-4o, the framework responds to the prompt ``Generate a visualization of a diver submerged in the ocean, with bubbles advected along the computed streamlines of the flow field.", yielding a realistic synthesized image reflecting the flow features obtained from the CFD simulation, as shown in Fig.~\ref{man}(e).

\begin{figure}[htbp]
    \centering
    \begin{subfigure}{\textwidth}
        \centering
        \includegraphics[width=\textwidth]{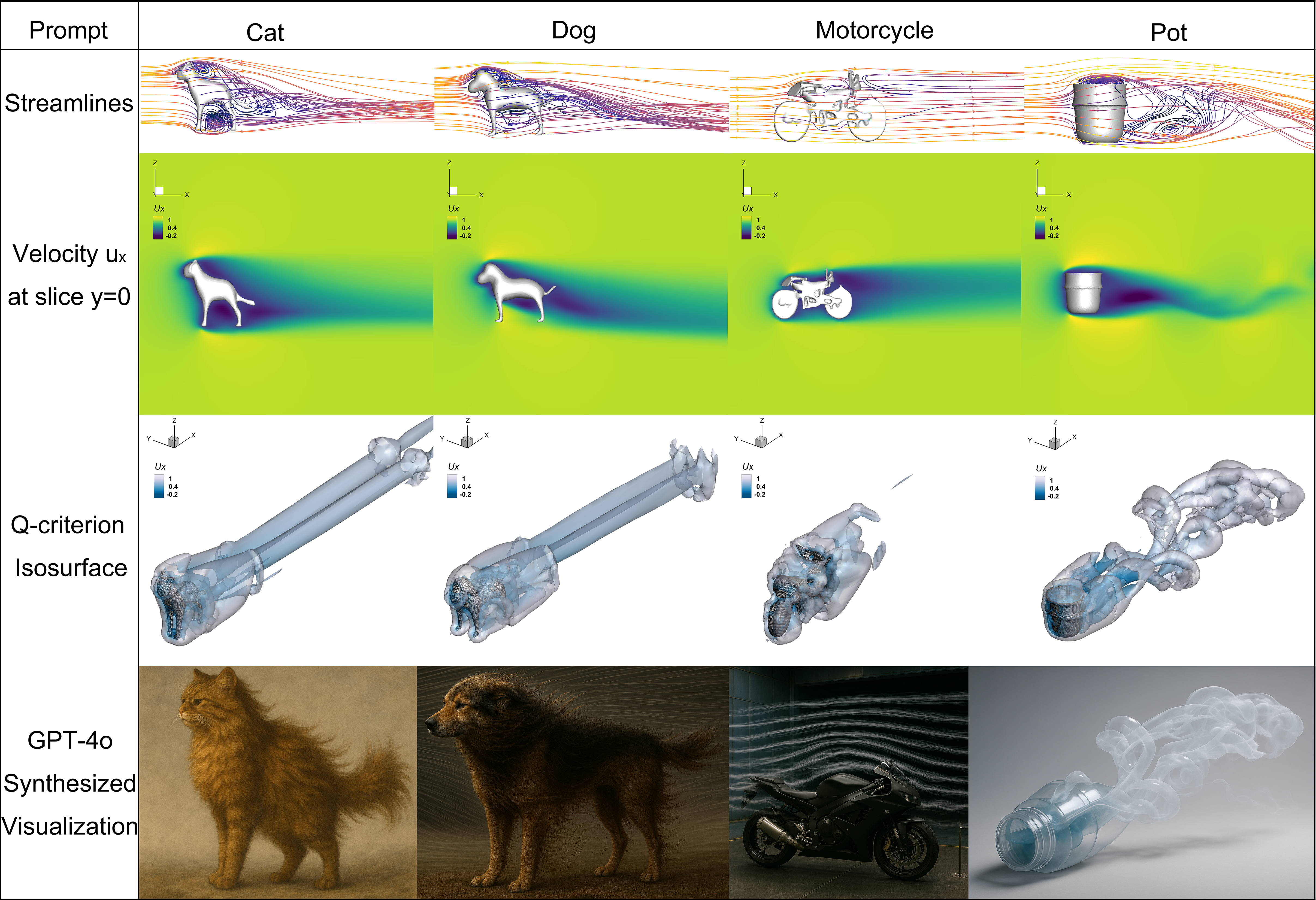}
    \end{subfigure}
    \begin{subfigure}{\textwidth}
        \centering
        \includegraphics[width=\textwidth]{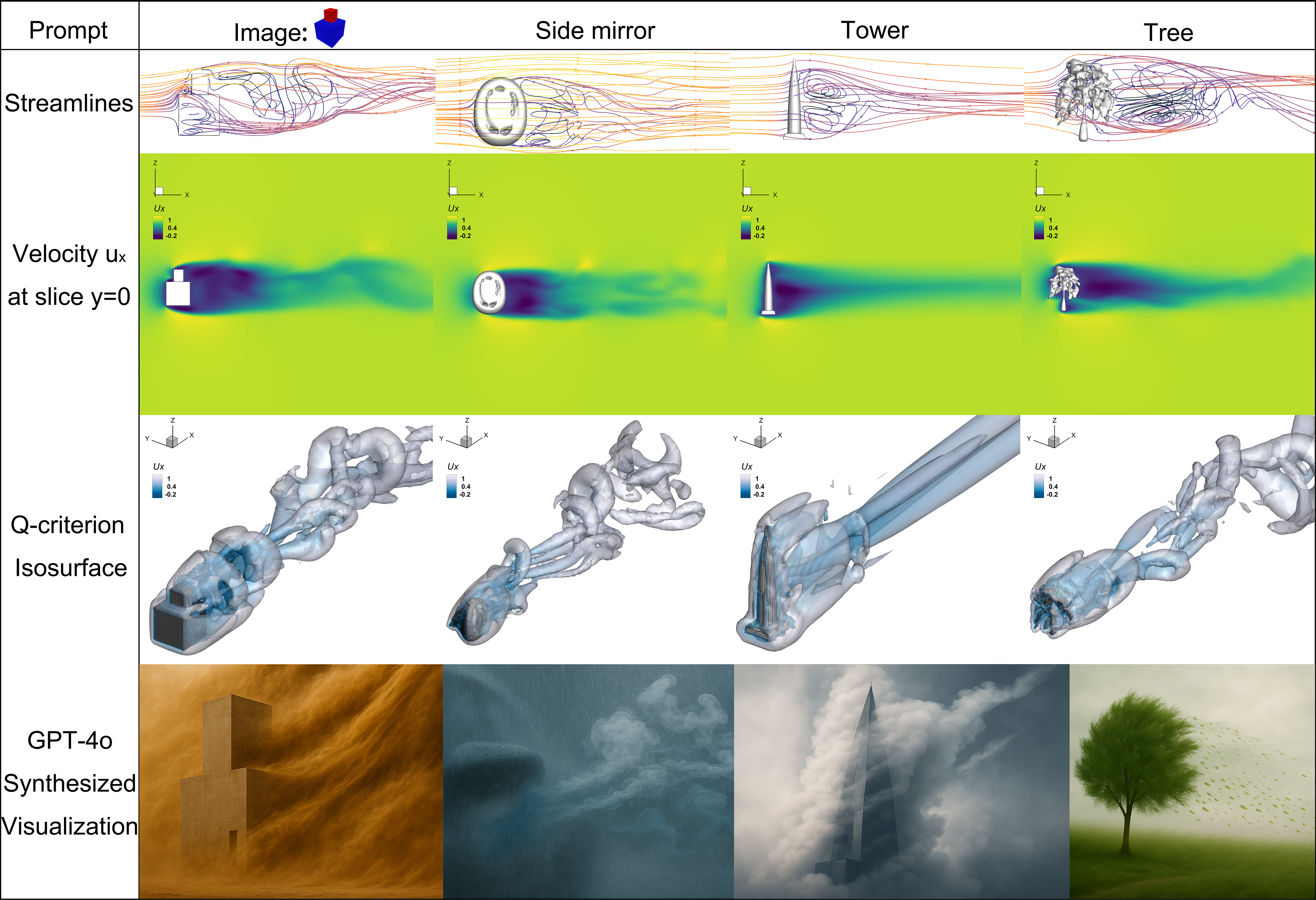}
    \end{subfigure}
    \caption{\textbf{CFDagent postprocessing results demonstrating generalization for different prompts:} Streamlines, velocity distributions, vortex structures ($Q$-criterion iso-surface), and GPT-4o synthesized visualizations.}
    \label{fig:generalization}
\end{figure}

To further validate the generalization capabilities of our proposed LLM-CFD framework, we examine simulations prompted by diverse textual inputs describing various complex geometries beyond the sphere example.
Specifically, geometries ``cat", ``dog", ``motorcycle", ``pot", ``side mirror", ``tower", and ``tree" are autonomously generated and simulated under identical conditions as the previous simulation of human geometry.
Fig.~\ref{fig:generalization} demonstrates representative results obtained for each case.
For each geometry, the velocity fields, streamlines, and vortex structures are systematically visualized, alongside GPT-4o synthesized images reflecting specific flow-related scenarios described by natural language prompts, which are consistent with the ``human" case.
Notably, the velocity distributions on the symmetry plane consistently reveal flow around sharp edges and stagnation points at frontal surfaces.
Streamline visualizations clearly illustrate distinct wake patterns, reflecting flow separation behavior specific to each geometry.
For example, the complex wake behind the ``pot" geometry exhibits dispersed vortical structures, contrasting sharply with the more streamlined wake structures observed behind the ``motorcycle" geometries.
The three-dimensional vortex structures, visualized via the $Q$-criterion iso-surfaces, emphasize differences in vortex shedding intensity and frequency across the geometries.
The ``side mirror" case, in particular, shows pronounced vortex shedding, indicative of complex, unsteady wake dynamics.
Finally, the framework demonstrates its multi-modal synthesis capabilities through GPT-4o-generated images, effectively visualizing realistic, scenario-specific phenomena inspired by the computed CFD results.

Collectively, these results underscore the robustness, versatility, and generalizability of the zero-shot CFDagent framework in autonomously simulating and visualizing complex flow fields around arbitrary geometries defined solely by natural language inputs.

\section{Conclusion}\label{sec13}
In this study, we introduce CFDagent, a zero-shot, natural-language-driven multi-agent framework capable of autonomously performing complex CFD simulations.
CFDagent integrates advanced LLM capabilities, a hybrid text-to-3D diffusion model, and an IB solver, enabling robust geometry generation and accurate flow simulation for arbitrary, user-defined geometries.
This approach significantly simplifies the CFD workflow, allowing users to initiate simulations and perform preliminary analyses without detailed expertise or extensive pre-processing.
CFDagent supports multi-mode inputs for geometry generation, including natural language descriptions of object shapes, 2D images, and 3D geometry meshes.
It also offers diverse output modalities, including quantitative plots, 2D/3D visualizations, and even synthesized realistic images generated from simulation results.
CFDagent demonstrates strong generalization to simulate flows around diverse and complex geometries, underscoring its promise as an intelligent and accessible assistant to simulate and interpret real-world flow phenomena.

Despite its current strengths, CFDagent exhibits some limitations that must be addressed to further enhance its practicality and efficiency. The limited quality and controllability of the generated geometries hinder precision.
Performance is constrained by the inherent complexity of CFD simulations, leading to significantly prolonged processing times.
Future improvements in AI-driven geometry generation quality, increased geometric controllability, and optimized computational strategies could significantly alleviate these constraints.

The architecture of CFDagent offers strong potential for further development, especially through integration with rapidly evolving AI techniques.
Notably, the IB solver's inherent capacity to handle dynamic meshes offers exciting potential for simulating flows around moving or deforming objects, expanding applicability of CFDagent to diverse, real-world scenarios such as biolocomotion or dynamic engineering components.
Additionally, coupling CFDagent with emerging generative AI models could facilitate the automatic production of high-quality, informative visualizations, supporting more comprehensive fluid dynamics analysis. 
Ultimately, CFDagent could be integrated with more powerful general-purpose language models to provide more intuitive guidance for simulations and deeper analysis of results.

\section{Methods}\label{sec:method}
\subsection{Zero-shot multi-agent system}\label{subsec:pipeline}
The zero-shot multi-agent system integrates natural language interaction with three specialized agents—the Preprocessing Agent, the Solver Agent, and the Postprocessing Agent—to autonomously configure and execute CFD simulations from natural language instructions. 
Implemented as an interactive web application using Streamlit~\citep{streamlit}, the framework incorporates an OpenAI GPT-based assistant~\citep{GPT4} to facilitate user-agent dialogue and task coordination.
Through the three agents, the user’s high-level natural language input is transformed into a fully configured simulation, with minimal manual effort required, as described in the algorithm shown in Fig.~\ref{fig:Workflow} and detailed in Supplementary Information.

The Preprocessing Agent initiates dialogue and determines whether the user intends to run a fluid-dynamics simulation and generate the corresponding object.
In the first stage, the user engages with a GPT-driven conversational interface.
If the user expresses interest in a fluid simulation, GPT will start to assist in simulation.
Through a dialogue, the user is asked to specify the object of interest either by name or description (e.g., ``dog" or ``sphere"), by uploading a reference image of the object, or by providing a pre-existing 3D mesh file.
The user also specifies the flow conditions, notably the desired Reynolds number for the simulation—a dimensionless quantity that characterizes the nature of the flow.
If the object is provided only as a text description or an image, the framework generates a corresponding 3D model along with its surface mesh.
This is accomplished using OpenAI’s Point-E, which can produce a three-dimensional point cloud representation of an object from either a textual description or a single image.
Once a sufficient point cloud is obtained, a signed distance function (SDF) regression model is applied to convert the point cloud into a continuous volumetric representation, and a marching cubes algorithm extracts a watertight surface mesh from this field.
The resulting triangular mesh, also referred to as the Lagrangian mesh, approximates the target object’s geometry.
If the user instead uploads an existing 3D mesh in the Preprocessing Agent, the generative modeling step is skipped and the provided geometry is used directly.
Through an agent-provided interface, the user defines the incoming flow direction with respect to the object, ensuring correct geometric orientation within the simulation.
At the end, the geometry mesh is finalized and properly positioned.

Within the Solver Agent, the simulation parameters are first configured, followed by the launching of the simulation.
First the user is prompted to provide essential CFD settings: the total simulation duration (end time), the Courant–Friedrichs–Lewy (CFL) number for numerical stability, the time step, and the output frequency for saving simulation data.
The GPT-based assistant suggests default or recommended values for these inputs.
Utilizing GPT, users are able to ask questions regarding any unfamiliar or unclear physical parameters at any stage.
The Eulerian grid has a default configuration, but a new mesh can also be specified by setting the grid size and computational domain, as is illustrated in Supplementary Information.
With the geometry and simulation parameters specified, the framework automatically generates the necessary configuration files for the CFD solver.
In particular, it writes out a Para.txt file containing the solver settings and global parameters, and a Bc\_Ic.txt file specifying the boundary and initial conditions as well as the Eulerian grids file.
Once all the necessary files are properly prepared, the solver proceeds to simulate the flow around the object using the IB method with the given mesh and parameters. 
A detailed description of the IB method is provided in section~\ref{subsubsec:ib}.

Within the Postprocessing Agent, we perform postprocessing to visualize the simulation results and extract quantitative metrics for analysis.
All postprocessing is performed collaboratively by the GPT-assisted agent and custom scripts: GPT first adjusts script parameters based on the user prompt, then executes the scripts accordingly.
The framework explicitly enables automated extraction and visualization of probe-based time-series data, contour plots from cross-sectional slices, three-dimensional rendering of iso-surfaces, and lift/drag coefficient histories.
\begin{figure}[h]
\centering
\includegraphics[width=1.0
\textwidth]{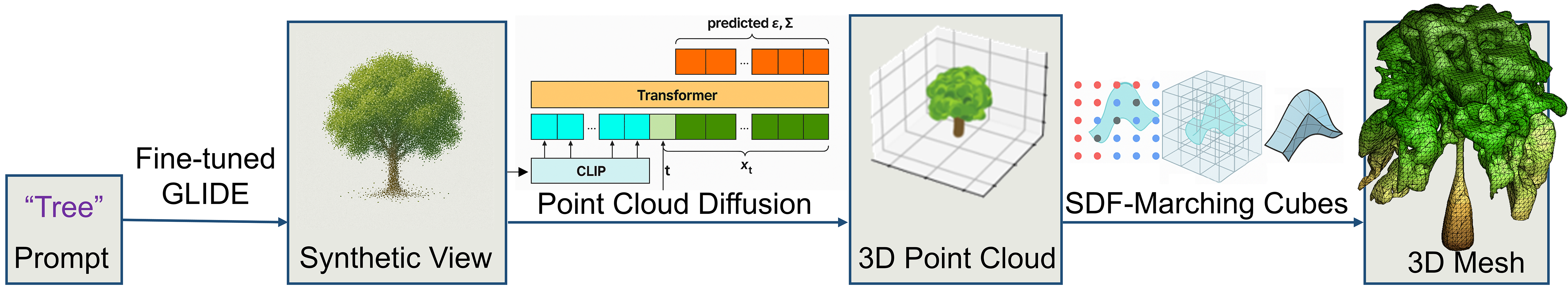}
\caption{\textbf{Text-to-3D Mesh Generation Pipeline.} A natural language prompt is processed to a fine-tuned GLIDE model to generate a synthetic image, which is then converted into a 3D point cloud using point cloud diffusion with CLIP and a transformer. The resulting point cloud is transformed into a 3D mesh using SDF estimation and Marching Cubes.}\label{meshGeneration}
\end{figure}

\subsection{Relevant technologies}\label{subsec:core-tech}
\subsubsection{Large language models}\label{subsubsec:llm}
LLMs such as GPT-4 series~\cite{GPT4} have demonstrated strong zero-shot reasoning and multimodal understanding capabilities.
In the context of CFD processing, these strengths translate into (i)~robust intent recognition from noisy user queries, (ii)~automatic extraction and validation of simulation parameters, and (iii)~context-aware tutoring for novice users. 
Unlike traditional rule-based chatbots, LLMs can generalize to unforeseen phrasing while enforcing domain-specific constraints through carefully designed prompts.

Using strictly formatted command blocks within a three‑tier slot‑filling structure—system message, command block, and self‑check—ensures deterministic parsing and natural dialogue, with any missing slot prompting exactly one concise follow‑up question.
After benchmarking several candidates on manually curated conversational test cases, we selected \textbf{GPT-4o} for production and retained \textbf{GPT-4o-mini} as a fallback.  
GPT-4o-mini is more cost-effective than GPT-4o, maintaining a pass@1 accuracy of $\ge 91\%$ on our task-oriented test suite.
Switching the production model only requires editing a single line in the Streamlit service, minimizing deployment friction.

\subsubsection{3D geometry generation based on Point-E}\label{subsubsec:pointe}
In this study, we employ the Point-E framework~\cite{nichol2022pointegenerating3dpoint} to efficiently generate 3D geometries from text prompts or 2D images.
As illustrated in Fig.~\ref{meshGeneration}, the pipeline involves generating a synthetic image from a text prompt using a fine-tuned GLIDE model, converting it into a 3D point cloud, and then into a 3D mesh.
The key idea of Point-E is to decouple the complex task of text-to-3D mapping into two successive diffusion-based generative steps: (1) text-to-image generation, and (2) image-to-point-cloud synthesis.
If a 2D image is provided, the framework directly proceeds to the second stage.

The first stage of our pipeline involves generating a synthetic rendered view $I$ from a given text prompt using a pretrained text-to-image diffusion model based on Guided Language-to-Image Diffusion for Generation and Editing (GLIDE)~\citep{nichol2021improveddenoisingdiffusionprobabilistic}.
The GLIDE model builds on the Denoising Diffusion Probabilistic Model (DDPM) framework~\cite{ho2020ddpm}, which defines a forward process that progressively corrupts data $q(\mathbf{x}_0)$ by incrementally adding Gaussian noise across $T$ timesteps, producing latent variables $\mathbf{x}_1, \mathbf{x}_2, \ldots, \mathbf{x}_T$.
At each step $t$, the conditional distribution is defined as
\begin{equation}
q(\mathbf{x}_t \mid \mathbf{x}_{t-1}) = \mathcal{N}\left( \mathbf{x}_t; \sqrt{1-\beta_t}\mathbf{x}_{t-1}, \beta_t \mathbf{I} \right),
\end{equation}
where $\beta_t \in (0, 1)$ is a small positive scalar from a predefined variance schedule that determines the noise magnitude injected into $\mathbf{x}_{t-1}$ at step $t$.
Intuitively, as $t \to T$, the data is progressively transformed into isotropic Gaussian noise. 
A model is then trained to approximate the reverse denoising process, enabling the generation of data.
More details can be found in the Supplementary Information.
The GLIDE model, which extends the standard Denoising Diffusion Probabilistic Model (DDPM) framework by introducing text conditioning, generates the rendered view $I$ from a given text prompt
\begin{equation}
I \sim p_{\theta_1}(\,\cdot\mid \mathrm{text}),
\end{equation}
where $\theta_1$ represents the pretrained parameters of the GLIDE model.
Crucially, the GLIDE model is fine-tuned on a dataset of 3D renderings, so that the synthesized images are biased towards plausible single-view projections of 3D objects.

The second stage involves generating a 3D point cloud $\mathcal{P}$ conditioned on the synthesized image $I$ via a point cloud diffusion model with the pretrained parameters $\theta_2$~\cite{nichol2022pointegenerating3dpoint}
\begin{equation}
\mathcal{P} \sim p_{\theta_2}(\cdot \mid I), \quad
\mathcal{P} = \{(\mathbf{p}_i, \mathbf{c}_i)\}_{i=1}^N,\quad
\mathbf{p}_i \in \mathbb{R}^3,\quad
\mathbf{c}_i \in [0,1]^3.
\end{equation}
where $\mathbf{p}_i$ denotes the 3D coordinates of the $i$-th point, and $\mathbf{c}_i$ represents RGB color.
The model directly outputs a tensor of shape $N\times6$.
To condition on the image, a frozen, pretrained CLIP model encodes the synthetic view $I$ into a spatial feature grid, which is projected into transformer tokens serving as conditioning input.
The noised point cloud $\mathbf{x}_t$ at timestep $t$, alongside a learned embedding of $t$, are tokenized and fed into the transformer.
The transformer model then predicts both the mean and variance of the reverse distribution:
\begin{equation}
p_{\theta_2}(\mathbf{x}_{t-1} \mid \mathbf{x}_t,I) = \mathcal{N}(\mu_{\theta_2}( \mathbf{x}_t,t, I), \Sigma_{\theta_2}(\mathbf{x}_t,t, I)),
\end{equation}
where output tokens corresponding to $\mathbf{x}_t$ are used to compute $\mu_\theta$ and $\Sigma_\theta$, typically by predicting the noise $\mathbf{\epsilon}$ and using DDPM parameterizations~\citep{ho2020ddpm}.
In practice, the stage employs a hierarchical two-step procedure: the point cloud diffusion model produces a coarse point cloud with $1024$ points, and an upsampler model refines it to $4096$ points. 

For simulations, we convert the generated point cloud $\mathcal{P}$ into a watertight triangular mesh by learning an implicit representation via a signed distance function (SDF).
A transformer-based network $\phi_\psi:\mathbb{R}^3 \rightarrow \mathbb{R}$ takes arbitrary 3D spatial queries $\mathbf{x} \in \mathbb{R}^3$ as input to predict their signed distance relative to the object's surface, where
$\phi_\psi(\mathbf x)<0$ denotes points outside the surface and $\phi_\psi(\mathbf x)>0$ denotes points inside.
The object's surface is defined as the zero-level set $\mathcal S=\{\mathbf x\in\mathbb R^3 \mid \phi_\psi(\mathbf x)=0\}$, which is subsequently converted into a watertight triangular mesh with Marching Cubes, followed by Laplacian smoothing to attenuate voxelization artifacts. 

\subsubsection{Immersed boundary method for incompressible flows}\label{subsubsec:ib}

The IB method offers an efficient framework for simulating incompressible flows around bodies with complex geometries on non-conformal, typically Cartesian, grids. This is achieved by introducing a forcing into the Navier–Stokes equations to enforce the boundary conditions at the fluid–body interface. In this formulation, the fluid domain is discretized on a fixed Eulerian grid, while the immersed boundaries are represented by Lagrangian markers that are either stationary or in motion.

For the momentum, the Navier–Stokes equations with IB forcing take the form:

\begin{align}
   \frac{\partial \mathbf{u}}{\partial t}+\mathbf{u}\!\cdot\!\nabla\mathbf{u}&=-\nabla p+\frac{1}{Re}\nabla^{2}\mathbf{u}+\mathbf{f}\\
   \nabla\!\cdot\!\mathbf{u}&=0,
 \end{align} 
where $\mathbf{u}(\mathbf{x},t)$ is the Eulerian velocity field, $p(\mathbf{x},t)$ the pressure, and $Re$ the Reynolds number. The term $\mathbf{f}$ represents the body force exerted on the fluid by the immersed surface. Within the IB framework, one defines a Lagrangian force $\mathbf{F}(\mathbf{X},t)$ on the fluid–body interface $S$ (with Lagrangian coordinate $\mathbf{X}$). This force is spread to the Eulerian grid via a convolution with a regularized delta function $\delta_h$:
\begin{equation}
    \mathbf{f}(\mathbf{x},t) = \int_{\mathbf{X}\in S} \mathbf{F}(\mathbf{X}, t)\, \delta_h(\mathbf{x} - \mathbf{X})\, d \mathbf{X}.
\end{equation}
Here $\delta_h$ ensures a smooth transfer of force from the Lagrangian markers to the Eulerian grid. Conversely, the fluid velocity at the Lagrangian markers is obtained by interpolating the Eulerian field: $\mathbf{U}^{*}(\mathbf{X},t) = \int_{\mathbf{x}\in \Omega} \mathbf{u}(\mathbf{x},t) \delta_h(\mathbf{x}-\mathbf{X}) d\mathbf{x}$.   
The Lagrangian force $\mathbf{F}$ is then chosen so as to enforce the no‐slip and no‐penetration conditions. In practice, one computes $\mathbf{F}(\mathbf{X},t)$ to drive the interpolated velocity $\mathbf{U}^*(\mathbf{X},t)$ toward the prescribed boundary velocity $\mathbf{U}_b(\mathbf{X},t)$ (for example, via a penalty or direct‐forcing formulation), thereby closing the coupling between fluid and body.

The spatial discretization is performed using a second-order finite volume scheme, and temporal integration is carried out with a three-step, second-order, low-storage Runge–Kutta method.
To address large-scale three-dimensional simulations efficiently, the solver employs parallelization based on domain decomposition and MPI protocols.
Flow variables are distributed across worker processors, whereas the IB equations are assembled and solved centrally on the root processor using a gather–scatter communication strategy, thus demonstrating excellent strong scalability.
More numerical details and validations can be found in Supplementary Information and our previous publications~\citep{WANG20113479}.

\backmatter








\section*{Data availability}
All the simulations carried out in this study can be found under DOI:10.5281/zenodo.15383155. 
Source data are provided with this paper.

\section*{Code availability}
An open-source version of the CFDagent framework has been released at DOI:10.5281/zenodo.1538315 as well. Access to the proprietary GPT-4 API can be obtained through OpenAI.










\begin{appendices}






\end{appendices}


\bibliography{sn-bibliography}
\section*{Acknowledgements}

This work was supported by the NSFC Excellence Research Group Program for `Multiscale Problems in Nonlinear Mechanics’ (No. 12588201), the National Natural Science Foundation of China (Nos. 12425207 and 92252203), the Chinese Academy of Sciences Project for Young Scientists in Basic Research (Grant No. YSBR-087). The work was also supported by the OCTAVE project (Grant No. P2024-01011) in the VINNOVA Strategic Vehicle Research and Innovation Program from the Swedish Energy Agency. 
\section*{Author contributions}
Z.X., G.H, S.W. and H.-D.Y. conceived the work. Z.X., S.W. and H.-D.Y. performed the system development, numerical simulations and results analysis. Z.X., L.W., Q.L., C.W., Y.C. and H.-D.Y. wrote the manuscript. 
All authors contributed to the analysis of the results. 
\section*{Competing interests}
The authors declare no competing interests.
\end{document}